\documentclass[a4paper, 11pt]{article}
\usepackage[affil-it]{authblk} 
\usepackage{etoolbox}
\usepackage{lmodern}
\usepackage[a4paper]{geometry}
\usepackage[in]{fullpage}
\usepackage[utf8]{inputenc}
\usepackage{amsmath}
\usepackage{amssymb}
\usepackage{cite}
\usepackage{amsfonts}
\usepackage{tensor}
\usepackage{footnote}
\usepackage[dvipsnames]{xcolor}
\usepackage[colorlinks]{hyperref}
\usepackage{subfiles}

\numberwithin{equation}{section}
\hypersetup{linkcolor=blue,citecolor=blue,urlcolor=blue}

\DeclareMathAlphabet{\mathpzc}{OT1}{pzc}{m}{it}
\DeclareMathAlphabet{\mathcalligra}{T1}{calligra}{m}{n}

\usepackage[framemethod=TikZ]{mdframed}
\mdfdefinestyle{MyFrame2}{
	linecolor=blue!20,
	middlelinewidth=2pt,
	frametitlerule=true,
	frametitlebackgroundcolor=blue!20,
	frametitlerulecolor=blue!20,
	frametitlerulewidth=1pt,
	innertopmargin=\topskip}

\title{Nonrelativistic superfluids in cosmology from a relativistic approach: Revisiting two formulations of superfluidity}

\makeatletter
\patchcmd{\@maketitle}{\LARGE \@title}{\fontsize{16}{19.2}\selectfont\@title}{}{}
\makeatother

\author[1]{Aline Favero\footnote{Email: \href{mailto:aline.favero@mail.mcgill.ca}{aline.favero@mail.mcgill.ca}}}

\author[1,2]{Heliudson Bernardo\footnote{Email: \href{mailto:heliudson_bernardo@brown.edu}{heliudson\_bernardo@brown.edu}}}

\affil[1]{Department of Physics, McGill University, Montreal, QC, H3A 2T8, Canada}

\affil[2]{Brown Theoretical Physics Center and Department of Physics, Brown University, \protect\\
Barus Building, 340 Brook Street, Providence, RI 02912, USA}

\date{\vspace{-5ex}}

\begin{document}

\maketitle

%%%%%%%%%%%%%%%%%%%%%%%%%%%%%%%%%%%%%%%%%%%%%%%%%%%%%%%%%%%%%%%%%%%%%%%%%%%%%%%%%%%%%%%%%%%%%%
\begin{abstract}
Two formulations of superfluidity are reviewed: Landau's phenomenological two-fluid model and a relativistic effective field theory description. We demonstrate how the two-fluid formalism can be recovered from the nonrelativistic limit of the relativistic effective theory at finite temperatures. We show how self-gravitating, nonrelativistic superfluids are obtained from the Newtonian limit of the relativistic approach on curved spaces. The concepts are presented in an accessible manner for readers who may not be deeply familiar with superfluidity from a condensed matter perspective.
\end{abstract}

%%%%%%%%%%%%%%%%%%%%%%%%%%%%%%%%%%%%%%%%%%%%%%%%%%%%%%%%%%%%%%%%%%%%%%%%%%%%%%%%%%%%%%%%%%%%%%
\section{Introduction} 
\label{sec:intro}

Despite the series of successful predictions brought forth by the Lambda-Cold Dark Matter ($\Lambda$CDM) model \cite{Lahav:2024npe,refId0,hoekstra2004properties,bennett2013nine,zwicky1933redshift,1936ApJ....83...23S,1970ApJ...159..379R,1970ApJ...160..811F,1973ApJ...186..467O,Anderson_2014,Tegmark_2004}, the fundamental nature of Dark Matter (DM) remains unknown. On large scales, the standard cosmological model describes DM as a cold, pressureless fluid of non-interacting particles, and while this description suffices for explaining large-scale observations, there are several discrepancies to be found in galactic scales (for a recent overview of the small-scale problems, see \cite{Bullock:2017xww, weinberg2015cold}, and \cite{751a,Colin_2000} for alternative approaches).

These discrepancies can be explained by the Superfluid Dark Matter (SFDM) model \cite{Berezhiani:2015bqa,Berezhiani:2015pia,2015arXiv150703013K} (see also \cite{Sin:1992bg,Goodman:2000tg,Peebles:2000yy,Hu:2000ke,Silverman:2002qx,Boehmer:2007um,Chavanis:2011zi,Bettoni:2013zma,Guth:2014hsa,Hui:2016ltb,Ferreira:2018wup,Das:2014agf,Das:2018udn, Das:2022mgr} and the reviews \cite{Ferreira:2020fam,Hui:2021tkt,Khoury:2021tvy}). In this model, DM consists of ultralight particles with a mass on the order of $10^{- 20}$ to a few eV \cite{Berezhiani:2015bqa}. On cosmological scales, DM still behaves as a cold, collisionless fluid, similar to its behaviour on the standard model; on galactic scales, however, the DM particles can thermalize and condense into a superfluid phase. This occurs in regions of high DM density, such as the core of galaxies, where the density surpasses a critical threshold, triggering a phase transition and causing the particles to form a Bose-Einstein Condensate (BEC). Additionally, the superfluid phonons couple to ordinary matter, mediating a long-range force between baryons that, for a specific superfluid equation of state, reproduces the MOdified Newtonian Dynamics (MOND) law \cite{Berezhiani:2017tth}.

At finite temperature, the phenomenology of superfluidity is described by Landau's two-fluid model \cite{Landau,landau1941j}, originally developed to describe superfluid helium. Below a critical temperature $T_{c}$, the superfluid is treated as a mixture of two components: a superfluid and a normal component. The relative concentrations of these components depend on the temperature: at $T = 0$, the system consists entirely of the superfluid component, but as the temperature rises, the proportion of the superfluid decreases while the normal component fraction grows. For $T \geq T_{c}$, the superfluid component vanishes entirely, leaving only the normal component.

While a nonrelativistic approach can be appropriate for studying low-energy systems such as liquid helium, studying high-energy superfluids requires a fully relativistic framework \cite{schmitt2015introduction}. Relativistic effects become significant when the particle's kinetic energy is comparable to their masses, and given the vast range of energy scales over which superfluidity is observed, a relativistic treatment becomes essential. For instance, the critical temperature of different systems can range from as low as $10^{-7}$~K in ultra-cold atomic gases to $10^{8}$~K for neutron matter, as indicated by astrophysical observations \cite{Page:2010aw} (see \cite{Rajagopal:2000wf} for a review on nuclear matter at high densities and the formation of relativistic superfluids in that context). Moreover, many UV descriptions for the origin of the superfluid DM are fully relativistic, e.g. see \cite{Witten:1984rs, Farhi:1984qu,Zhitnitsky:2002qa, Alexander:2018fjp, Alexander:2020wpm, Ouyed:2023hqe, Alexander:2024qml}, and the relativistic description might be more efficient when the nonrelativistic fluid is considered in a relativistic background (e.g. see \cite{Bernardo:2023ehz} for a study on the behaviour of superfluid DM on a cosmic string background). In addition to the importance of relativistic effects, the relativistic approach is also more general: the nonrelativistic case can be recovered as a limiting case.

Although there are equivalent formulations of relativistic superfluids \cite{ISRAEL1981, KHALATNIKOV1982,lebedev1982relativistic,ISRAEL1982_2,dixon1982thermodynamics,Carter:1992gmy,Carter:1995if,Son:2000ht}, this work considers the operational approach through the low-energy effective field theory framework proposed by Nicolis \cite{Nicolis:2011cs}. This framework provides a fully relativistic and Poincar\'e invariant description of finite temperature superfluids, in which the dynamics come straightforwardly from a low-energy effective action via the standard variational principle. The action is derived in the traditional effective field theory approach: the infrared degrees of freedom are identified -- the superfluid phase and the comoving coordinates of the volume elements of the normal fluid --, and the most general action involving these degrees of freedom is constructed, ensuring that it remains consistent with the symmetries of the system. To the best of our knowledge, there is no explicit construction of the nonrelativistic limit of such formulation, which presumably would reproduce the Galilean description of superfluids usually assumed in cosmology.

In this paper, we show how Landau's two-fluid model is recovered from the relativistic effective description. We revisit both the two-fluid formalism and the effective field theory approach to superfluidity to create a map between the field variables and the standard hydrodynamical and thermodynamical ones in the nonrelativistic limit. This paper is intended for cosmologists and astrophysicists who work with superfluid dark matter but may have little familiarity with the details of superfluidity in condensed matter physics. We aim to present these concepts in a way that is accessible without requiring deep prior knowledge of superfluidity from a condensed matter perspective. We provide a bridge between the general relativistic effective field description of superfluid and the self-gravitating, nonrelativistic description typically employed in cosmology. In doing so, we hope to close a literature gap on explicitly recovering the nonrelativistic limit from general symmetry principles, including Lorentz invariance.

This paper is structured as follows: section \ref{sec:tb} establishes the theoretical background, with a brief overview of Landau's two-fluid model in section \ref{ssec:tf} and of the effective theory of superfluidity in section \ref{ssec:eft}, defining their main equations and variables; in section \ref{sec:nr}, we recover the nonrelativistic limit from the effective theory in different regimes and discuss how to obtain self-gravitating superfluids (which are typically used in cosmology) from the Newtonian limit of the relativistic theory. In section \ref{sec:map}, we establish a map between the two-fluid model and the nonrelativistic limit of the field theory approach at finite temperature and finally, in section \ref{sec:conc}, we conclude. Throughout this paper, we use natural units in which the speed of light, Boltzmann's and Planck's constant are set to one, i.e. $c = k_{B} = \hbar = 1$. We adopt the metric signature $(-,+,+,+)$.

\section{Preliminaries}
\label{sec:tb}

To set the notation and the background for the rest of the paper, in this section, we briefly review the two-fluid model for nonrelativistic superfluids \cite{landau2013fluid} and the effective field approach for relativistic superfluids~\cite{Nicolis:2011cs}.

\subsection{Two-fluid Model}
\label{ssec:tf}

The two-fluid model of superfluidity, first introduced by London \cite{London,london1938lambda} and Tisza \cite{tisza1938transport} in 1938 and later reformulated by Landau \cite{Landau,landau1941j}, in 1941 (see also \cite{Donnelly, BALIBAR2017586} for a more detailed account of the history of superfluid physics), describes the superfluid as a mixture of two different fluids: a superfluid component, composed of atoms in a BEC, and a normal component, made up of collective excitations. The former behaves as an inviscid fluid, carries zero entropy and flows irrotationally, while the latter behaves as an ordinary viscous fluid. The normal component is formed by two different types of quasi-particles, phonons -- lower-energy excitations -- and rotons -- higher-energy excitations. These two components are in relative motion, and we denote the velocity of the superfluid and normal components by $\vec{v}_{s}$ and $\vec{v}_{n}$, respectively.

In this framework, a superfluid with total particle number density $n$ will satisfy
\begin{align}
    n = n_{n} + n_{s},
\end{align}
where $n_{n}$ and $n_{s}$ give the contributions of the normal and superfluid components, respectively. Both $n_{n}$ and $n_{s}$ are temperature dependent: at $T = 0$, $n_{s} = n$ and $n_{n} = 0$, so that all the fluid is found in the superfluid state; as the temperature increases, excitations start to appear in the form of the normal component, and at $T = T_{c}$, all the superfluid component is gone, so $n_{s} = 0$ and $n_{n} = n$.

The mass density of the superfluid and normal parts are defined in terms of the particle number densities as
\begin{align}
    \rho_{n,m} &= m n_{n},  &   \rho_{s,m} &= m n_{s},
\end{align}
respectively, so that the total mass density is given by
\begin{align}
    \rho_{m} = \rho_{n,m} + \rho_{s,m}.
\end{align}

The mass flux density of the system, $\vec{j}$, or momentum density, is given by
\begin{align}
    \vec{j} = \rho_{s,m} \vec{v}_{s} + \rho_{n,m} \vec{v}_{n},\label{eq:e09}
\end{align}
i.e. a sum of the fluxes pertaining to the normal and superfluid parts.

With these quantities in hand, we can now summarize the equations of superfluid dynamics\cite{landau2013fluid}. The mass density $\rho_{m}$ and the flux $\vec{j}$ obey the continuity equation,
\begin{align}
    \frac{\partial\rho_{m}}{\partial t} + \vec{\nabla} \cdot \vec{j} = 0,\label{eq:e04}
\end{align}
which expresses the law of mass conservation.

The conservation of momentum is given by the equation
\begin{align}
    \frac{\partial j^{i}}{\partial t} + \frac{\partial \Pi^{ik}}{\partial x^{k}} = 0,\label{eq:e05}
\end{align}
where $\Pi^{ik}$ is the momentum flux density tensor, defined by
\begin{align}
    \Pi^{ij} = \rho_{n,m} v_{n}^{i} v_{n}^{j} + \rho_{s,m} v_{s}^{i} v_{s}^{j} + P \delta^{ij},\label{eq:e11}
\end{align}
with $P = P_{n} + P_{s}$ denoting the total pressure of the system.

The law of entropy conservation is given by
\begin{align}
    \frac{\partial\left(\rho_{m} s\right)}{\partial t} + \vec{\nabla} \cdot \left(\rho_{m} s \vec{v}_{n}\right) = 0,\label{eq:e08}
\end{align}
where the second term in parenthesis, $\rho_{m} s \vec{v}_{n}$, is the entropy flux.

Finally, the conservation of energy is given by the expression
\begin{align}
    \frac{\partial \epsilon}{\partial t} + \vec{\nabla} \cdot \vec{Q} = 0,\label{eq:e10}
\end{align}
where $\epsilon$ is the energy per unit volume of the fluid and $\vec{Q}$ the energy flux density vector, given by
\begin{align}
    \epsilon &= \epsilon_{n} + \epsilon_{s} + \frac{1}{2} \rho_{n,m} v_{n}^{2} + \frac{1}{2} \rho_{s,m} v_{s}^{2},\\
    \vec{Q} &= \vec{v}_{n} \left(\epsilon_{n} + \frac{1}{2} \rho_{n,m} v_{n}^{2} + P_{n}\right) + \vec{v}_{s} \left(\epsilon_{s} + \frac{1}{2} \rho_{s,m} v_{s}^{2} + P_{s}\right).
\end{align}
Here, $\epsilon_{n}$ and $\epsilon_{s}$ are the internal energy densities of the normal and superfluid components, respectively, which are given by the usual thermodynamic relations,
\begin{align}
    \epsilon_{n} &= \mu \rho_{n,m} + T s - P_{n},\\
    \epsilon_{s} &= \mu \rho_{s,m} - P_{m},
\end{align}
where $\mu$ is the chemical potential. Equations \eqref{eq:e04}, \eqref{eq:e05}, \eqref{eq:e08} and \eqref{eq:e10} provide the dynamics of nonrelativistic fluids. Our goal is to recover these from the general effective field approach for relativistic superfluids, which we revisit in the next section.

\subsection{Effective theory approach}
\label{ssec:eft}

From an effective field point of view, a BEC is described by a state that spontaneously breaks the global $U(1)$ symmetry associated with the conservation of particle number in the fluid. This is achieved by the condensation of excited particles into the ground state during the formation of the BEC. By the Goldstone theorem, there is a gapless excitation $\psi$ which non-linearly realizes the $U(1)$ symmetry as
\begin{align}
    \psi \rightarrow \psi + a, \quad \mathrm{where} \;a \;\mathrm{is \;a \;constant,}\label{eq:ss01}
\end{align}
and this is the only degree of freedom present at low energies.

The superfluid phase at $T = 0$ can be described by
\begin{align}
    \psi = \mu t,
\end{align}
where $\mu$ is the chemical potential. This homogeneous and isotropic state admits gapless excitations $\pi$,
\begin{align}
    \psi = \mu t + \pi,
\end{align}
the superfluid phonons. At finite temperature, the phonons will be excited, will reach thermodynamic equilibrium and form a thermal bath. Otherwise, if their frequencies are lower than the phonon inverse mean free time, and their wavelengths longer than their mean free path, this thermal bath of phonons will be describable by usual hydrodynamics. We are then left with a normal fluid made up of phonons which moves through and interacts with the background superfluid. As the temperature increases, it approaches a critical temperature $T_c$, above which the symmetry is restored and superfluidity is gone. Only the normal component remains, and the dynamics are once again those of an ordinary fluid.

We can thus postulate that, at temperatures between $0$ and $T_{c}$, our system will be composed of two fluids, a normal one and a superfluid one, and the density of the superfluid decreases monotonically with $T$, going to $0$ at $T_{c}$. This is the essence of Landau's two-fluid model.

For the rest of this section, we follow \cite{Nicolis:2011cs} closely. The low-energy degrees of freedom of our field theory are the superfluid phase $\psi\left(\mathbf{x}, t\right)$ -- which enjoys a shift symmetry --, and the comoving coordinates for the volume elements of the normal fluid component, which are parametrized by three scalar fields $ \phi^{i}\left(\mathbf{x}, t\right)$, $i = 1, 2, 3$. These $\phi^{i}$ are the comoving (or `Lagrangian') coordinates of the fluid element occupying physical (or `Eulerian') position $\mathbf{x}$ at time $t$. The fluid's internal space is homogeneous and isotropic, such that the dynamics of $\phi^{i}$ should obey~\cite{Dubovsky:2011sj,Andersson:2020phh}
\begin{align}
    \phi^{i} &\rightarrow \phi^{i} + a^{i}, &   &a^{i} = \mathrm{const.},\label{eq:ss02}\\
    \phi^{i} &\rightarrow R\indices{^{i}_{j}} \phi^{j}, &   &R \in SO(3).
\end{align}
Moreover, the $\phi^{i}$ should be invertible, $\partial_{i}\phi^{i} \neq 0$, and if the fluid is incompressible, volume preserving, $\det \partial_{j}\phi^{i} = 1$.

Lastly, we note that if we try to move around a volume element of a non-dissipative fluid, we would not feel any resistance; if we try to do the same with a solid, the motion would suffer resistance from the rest of the solid's volume. In other words, the fluid's dynamics must be invariant under volume-preserving reparametrizations \cite{Dubovsky:2011sj},
\begin{align}
    \phi^{i} \rightarrow \xi^{i}\left(\phi^{j}\right), \quad \det\frac{\partial \xi^{i}}{\partial \phi^{j}} = 1.\label{eq:vs01}
\end{align}

Given the shift symmetries \eqref{eq:ss01} and \eqref{eq:ss02}, $\psi$ and $\phi^{i}$ should enter the Lagrangian with at least one derivative acting on them, so at lowest order in the derivative expansion, $\mathcal{L}$ should be made up of $\partial_{\mu}\psi$ and $\partial_{\mu}\phi^{i}$. In order to obey the symmetry \eqref{eq:vs01}, the fluid variables have to appear through the combination
\begin{align}
   J^{\mu} &= \frac{1}{6} \epsilon^{\mu \alpha \beta \gamma} \epsilon_{ijk} \partial_{\alpha}\phi^{i}\partial_{\beta}\phi^{j}\partial_{\gamma}\phi^{k}.\label{eq:j}
\end{align}

Now we have the building blocks for our Lagrangian: $\psi$ should enter as $\partial_{\mu}\psi$, and $\phi^{i}$ as
$J^{\mu}$. With these vectors, we can construct three scalar quantities,
\begin{align}
    X &= \partial_{\mu}\psi \partial^{\mu}\psi,\\
    b &= \sqrt{- J^{\mu} J_{\mu}} = \sqrt{\det\left(\partial_{\mu}\phi^{i}\partial^{\mu}\phi^{j}\right)},\\
    y &= - \frac{1}{b} J^{\mu} \partial_{\mu}\psi \equiv u^{\mu} \partial_{\mu}\psi,\label{eq:y}
\end{align}
where $u^{\mu}$ is the normal fluid's four-velocity, since it's normalized to $-1$ and the fluid's comoving coordinates do not change along its integral curves, $u^{\mu} \partial_{\mu}\phi^{i} = 0$, and $\partial^{\mu}\psi$ can be interpreted as the four-velocity of the superfluid component (minus a normalization constant).

In summary, the low-energy Lagrangian density describing superfluidity has the form \cite{Nicolis:2011cs}
\begin{align}
    \mathcal{L} = F\left(X, b, y\right).
\end{align}
To compute the associated stress tensor, we vary $\mathcal{L}$ with respect to the metric, which gives
\begin{align}
    T^{\mu\nu} = \left(y F_{y} - b F_{b}\right) u^{\mu} u^{\nu} + \left(F - b F_{b}\right) g^{\mu\nu} - 2 F_{X} \partial^{\mu}\psi \partial^{\nu}\psi.\label{eq:tmn}
\end{align}
The current associated with the spontaneously broken $U(1)$ symmetry \eqref{eq:ss01} is just
\begin{align}
    N^{\mu} &= F_{y} u^{\mu} + 2 F_{X} \partial^{\mu}\psi,
\end{align}
the particle number four-current.

Now, we can match our field-theory quantities to thermodynamical ones. To do so, we need to pick a local reference frame, and since our system is composed of two fluids in relative motion, we can choose either component; we choose to move with the normal fluid component, along $u^{\mu}$.

The energy-momentum tensor for a perfect fluid has the general form
\begin{align}
    T^{\mu\nu} = \left(\rho + P\right) u^{\mu} u^{\nu} + P g^{\mu\nu},
\end{align}
from which we can isolate the energy density and pressure,
\begin{align}
    \rho &= T^{\mu\nu} u_{\mu} u_{\nu},\\
    P &= \frac{1}{3} T^{\mu\nu} \left(g_{\mu\nu} + u_{\mu} u_{\nu}\right).\label{eq:pres}
\end{align}
Replacing $T^{\mu\nu}$ in the above expression by \eqref{eq:tmn}, we can easily obtain the total energy density,
\begin{align}
    \rho &= T^{\mu\nu} u_{\mu} u_{\nu} = y F_{y} - F - 2 y^{2} F_{X}.
\end{align}
When computing the pressure, however, we have to remember that we are working with a system composed of two fluids, and that since there is no frame in which $T^{\mu\nu}$ is isotropic, equation \eqref{eq:pres} is not valid due to the relative motion of the fluid components. Instead, we have an energy-momentum tensor which is the sum of an isotropic tensor, $\propto g^{\mu\nu}$, and of two rank-one tensors, one $\propto u_{\mu} u_{\nu}$, and one $\propto \partial_{\mu}\psi \partial_{\nu}\psi$ -- which was earlier identified with the four-velocity of the superfluid component. Thus, we can identify the pressure with the coefficient of $g^{\mu\nu}$, i.e.
\begin{align}
    P = F - b F_{b}.
\end{align}

From the four-current $N^{\mu}$, we can extract the total particle number density,
\begin{align}
    n &= - N^{\mu} u_{\mu} = F_{y} - 2 y F_{X}.
\end{align}

Now that we have $\rho$, $P$ and $n$, we can use thermodynamic identities to find the remaining fluid variables. Starting from Euler's equation,
\begin{align}
    \rho + P = T s + \mu n,
\end{align}
and replacing in the expression the quantities we already have, we find
\begin{align}
    y \left(F_{y} - 2 y F_{X}\right) - b F_{b} = T s + \mu \left(F_{y} - 2 y F_{X}\right),
\end{align}
which is satisfied for
\begin{align}
    \mu &= y,   &   T s &= - b F_{b}.
\end{align}
In order to solve the ambiguity in the second expression, we take another look at the four-vector $J^{\mu}$: it follows from its definition \eqref{eq:j} that it is an identically conserved current, i.e. $\partial_{\mu}J^{\mu} = 0$; furthermore, $J^{\mu} = b u^{\mu}$, and we see that it is aligned with $u^{\mu}$. Thus, $J^{\mu}$ can be identified with the entropy four-current \cite{2012PhRvD..85h5029D},
\begin{align}
    J^{\mu} = s u^{\mu},
\end{align}
which leads to the identifications
\begin{align}
    s &= b,   &   T &= - F_{b}.\label{eq:e07}
\end{align}

Lastly, we want to express $X$ in terms of fluid variables. The vector $\partial_{\mu}\psi$ can be decomposed into the sum of two orthogonal vectors, one parallel to $u^{\mu}$ and one perpendicular:
\begin{align}
    \nonumber \partial_{\mu}\psi &= \left(\frac{u_{\alpha} \partial^{\alpha}\psi}{u_{\nu} u^{\nu}}\right) u_{\mu} + \xi_{\mu}\\
    &= - y u_{\mu} + \xi_{\mu}.
\end{align}
The component of $\partial_{\mu}\psi$ orthogonal to $u^{\mu}$, $\xi^{\mu}$, is just
\begin{align}
    \xi^{\mu} &= \partial^{\mu}\psi + y u^{\mu} = \left(\eta^{\mu\nu} + u^{\mu} u^{\nu}\right) \partial_{\nu}\psi,
\end{align}
which is a space-like vector\footnote{Since we are in a frame which is comoving to the normal component, we must have $u^{\mu} = \left(1, 0, 0, 0\right)$. Thus, because $\xi^{\mu} u_{\mu} = 0$, we know that $\xi^{\mu}$ must be space-like.}, $\xi^{\mu} = \left(0, \vec{\xi}\right)$. Since $X = \partial_{\mu}\psi \partial^{\mu}\psi$, we can write
\begin{align}
    X = - y^{2} + \xi^{2},\label{eq:e03}
\end{align}
where $\xi = \vert\vec{\xi}\vert$.

Computing the differential of \eqref{eq:pres}, we find
\begin{align}
    \nonumber dP &= F_{X} \:dX + F_{y} \:dy - b \:dF_{b}\\
    \nonumber &= s \:dT + n d\mu + F_{X} \left(dX + 2 y \:dy\right)\\
    &= s \:dT + n \:d\mu + 2 F_{X} \xi \:d\xi,
\end{align}
which defines the equation of state of the superfluid. Finally, from equation \eqref{eq:pres}, we can construct the Lagrangian
\begin{align}
    \nonumber \mathcal{L} &= P + b F_{b} = P - T s\\
    &= P - T \frac{\partial P}{\partial T}.
\end{align}

Analogously to what was done for $\partial_{\mu}\psi$, we can also decompose the superfluid component's four-velocity, $\tilde{u}^{\mu}$, into two orthogonal vectors, $u^{\mu}$ and $\xi^{\mu}$. The normalized four-velocity $\tilde{u}^{\mu}$ is given by
\begin{align}
    \tilde{u}^{\mu} = - \frac{\partial^{\mu}\psi}{\sqrt{- X}},
\end{align}
and it can be decomposed as
\begin{align}
    \nonumber \tilde{u}^{\mu} &= \left(\frac{u_{\nu} \tilde{u}^{\nu}}{u_{\alpha} u^{\alpha}}\right) u^{\mu} + \zeta^{\mu}\\
    &= \frac{y}{\sqrt{- X}} u^{\mu} + \zeta^{\mu},
\end{align}
where $\zeta^{\mu}$ is given by
\begin{align}
    \zeta^{\mu} = \tilde{u}^{\mu} - \frac{y}{\sqrt{- X}} u^{\mu}
    = - \frac{1}{\sqrt{- X}} \xi^{\mu}.
\end{align}

In the normal component's frame, $u^{\mu} = \left(1, \vec{0}\right)$ and $\xi^{\mu} = \left(0, \xi^{i}\right)$, so that
\begin{align}
    \tilde{u}^{\mu} &= \frac{y}{\sqrt{- X}} \left(1, - \frac{\xi^{i}}{y}\right) \equiv \gamma \left(1, \tilde{v}^{i}\right),
\end{align}
where $\tilde{v}^{i}$ is the velocity of the superfluid component, and $\gamma = \left(1 - \tilde{v}^{2}\right)^{-1/2}$ is the usual Lorentz gamma factor. When there is no relative velocity between the two components, $\xi^{\mu} = 0$, and $y = \sqrt{- X}$.

\section{Recovering the nonrelativistic limit for BEC DM}
\label{sec:nr}

In this section, we recover the nonrelativistic description of superfluid DM from the relativistic approach described in the previous section. We consider the low- and high-temperature limits. We also recover the self-gravitating case that is used in cosmology by assuming a generic curved metric and then taking the Newtonian limit.

\subsection{Low-temperature Limit}
\label{ssec:lt}

Let's suppose that $T=0$, such that the superfluid's normal component is absent and we can neglect the $b$ and $y$ dependence of $F$. In this case,
\begin{align}
    T_{\mu\nu} &= -2 F_X \partial_\mu \psi \partial_\nu \psi + F g_{\mu\nu},\\
    N^\mu &= 2 F_X \partial^\mu \psi,
\end{align}
and there is only the superfluid component frame. Note that
\begin{equation}
    \tilde{u}^\mu = -\frac{\partial^\mu \psi}{\sqrt{-X}}
\end{equation}
is a time-like, future-directed, and normalized four-vector that we can identify as the superfluid velocity. To check that, let's evaluate it at the homogeneous configuration $\psi = \mu t$:
\begin{equation}
    \tilde{u}^\mu= -g^{\mu\nu} \frac{\partial_\nu \psi}{\sqrt{-X}} = - g^{\mu\nu} \delta_\nu^0 \frac{\mu}{\sqrt{\mu^2}} = - g^{\mu 0}\text{sign}(\mu),
\end{equation}
and for Minkowski space, we have 
\begin{equation}
    \tilde{u}^\mu = \text{sign}(\mu)(1,0,0,0).
\end{equation}
So, if $\mu>0$, the four-velocity is future-directed. And we know that, in the nonrelativistic limit, the chemical potential $\mu$ is dominated by the rest mass $m$ of the superfluid particles, such that $\mu>0$. This justifies the minus sign in the definition of $\tilde{u}^\mu$ above.

Then, we can write 
\begin{subequations}
\begin{align}
    T_{\mu\nu} &= 2X F_X \tilde{u}_\mu \tilde{u}_\nu + Fg_{\mu\nu},\label{Tmunu}\\
    N^\mu &= -2\sqrt{-X}F_X \tilde{u}^\mu,\label{eq:e12}
\end{align}
\end{subequations}
and the energy density, pressure, and number density of the superfluid are, respectively,
\begin{align}
    \rho = 2X F_X - F, \quad   P = F, \quad n = - N^{\mu} \tilde{u}_\mu= -2\sqrt{-X} F_X.\label{eq:e13}
\end{align}

To get the nonrelativistic limit, we need to isolate the rest mass contribution to $\psi$:
\begin{equation}
    \psi = m t + \theta,
\end{equation}
where $\dot{\theta}\ll m$. In this case (and for flat space),
\begin{align}
    X &= -(m + \dot{\theta})^2 + (\nabla \theta)^2\nonumber\\
    &\approx -m^2 -2m Y,
\end{align}
where we defined 
\begin{equation}
    Y \equiv \dot{\theta}-\frac{1}{2m}(\nabla \theta)^2, 
\end{equation}
which is a typical variable for the Galilean effective approach to superfluids \cite{Khoury:2021tvy}. Note that, in the homogeneous case, $\dot{\theta}$ is the nonrelativistic chemical potential.

Moreover,
\begin{align}
    \nonumber \tilde{u}^\mu  &= \frac{(m+ \dot{\theta}, - \partial^i \theta)}{\sqrt{(m+\dot\theta)^2 - (\nabla\theta)^2}} = \frac{(1+ \dot{\theta}/m, - \partial^i \theta/m)}{\sqrt{(1+\dot\theta/m)^2 - (\nabla\theta)^2/m^2}}\\
    &\approx \left(1-\frac{Y}{m}\right)\left(1+ \frac{\dot{\theta}}{m}, -\frac{\partial^i\theta}{m}\right),
\end{align}
and in the nonrelativistic limit, the speed should be much smaller than unity, while the relativistic $\gamma$ factor is approximately one. This is precisely the case when $\dot \theta \ll m$ and $| \partial^i \theta|/m \ll 1 $, such that
\begin{equation}\label{nonrelspeed}
    \tilde{u}^\mu \approx \left(1, - \frac{\partial^i \theta}{m}\right) \equiv \left(1, \tilde{v}^{i}\right),
\end{equation}
and $Y$ can be written as
\begin{align}
    Y = \dot{\theta} - \frac{1}{2} m \tilde{v}^{2}.\label{eq:e06}
\end{align}

So, $F(X)$ should be seen as a function of $Y$ in the nonrelativistic limit:
\begin{equation}
    F(X(Y)) \equiv \mathcal{P}(Y), \quad F_X = \frac{dY}{dX} \mathcal{P}_Y = -\frac{\mathcal{P}_Y}{2m}.
\end{equation}
If the pressure $P = \mathcal{P}(Y)$ is much smaller than the energy density, we have
\begin{equation}
    \rho \approx 2X F_X \approx m \mathcal{P}_Y,
\end{equation}
while the number density is
\begin{equation}
    n \approx \mathcal{P}_Y.
\end{equation}
Thus, we find that $\rho \approx m n$ in the nonrelativistic limit, as expected.

The equation of motion for $\psi$, 
\begin{equation}
    \nabla_\mu (F_X \partial^\mu \psi) = -\nabla_\mu (\sqrt{-X}F_X \tilde{u}^{\mu}) = 0,
\end{equation}
simplifies to 
\begin{equation}
    \nabla_{\mu} (\mathcal{P}_{Y} \tilde{u}^\mu ) = 0,
\end{equation}
where $\tilde{u}^\mu$ is given by \eqref{nonrelspeed}. In flat space, this gives
\begin{equation}
    \frac{\partial n}{\partial t} - \frac{1}{m} \nabla \cdot (n \nabla \theta) = 0,
\end{equation}
which is just the continuity equation after multiplying by $m$ and identifying $nm = \rho$ and $v^{i} = - \partial^{i}\theta/m$:
\begin{equation}
    \frac{\partial \rho}{\partial t} + \nabla_{i} (\rho v^{i}) =0.
\end{equation}
Note that we could have also obtained that from the conservation of the energy-momentum tensor. This happens because, in the nonrelativistic limit, the conservation of mass and energy are related.

\subsection{Self-Gravitating superfluids}

Self-gravitating BEC can be described after assuming a non-trivial metric in the relativistic theory. Einstein's equations may be sourced by the energy-momentum tensor \eqref{Tmunu}, which has a rather complicated dependence on the metric because it enters in the definition of $X$. However, for weak gravity in the nonrelativistic limit, the equations can be greatly simplified; this is the Newtonian limit of the theory. Consider the linearized metric
\begin{equation}\label{newtonianmetric}
    ds^2 = -(1+2\Phi(x^i))dt^2 + (1-2\Psi(x^i)) \delta_{ij}dx^i dx^j,
\end{equation}
where the Newtonian potential $\Phi$ and curvature scalar $\Psi$ are much smaller than unity. If Einstein's equations are sourced by a perfect fluid, we find, after linearization,
\begin{subequations}
\begin{align}
    \nabla^2 \Phi &= 4\pi G (\rho + 3 P),\\
    \nabla^2 \Psi &= 4\pi G (\rho - P),
\end{align}
\end{subequations}
and if $P \ll \rho$, we have $\Phi = \Psi$. Moreover, energy and momentum conservation gives, to leading order in $v^i$,
\begin{subequations}
\begin{align}
    \dot \rho + v^i \partial_i\rho +(\rho + P)\partial_i v^i &= (\rho + P) v^i\partial_i (\Phi - 3 \Psi),\\
    (\rho + P) (\dot v^i + v^j \partial_j v^i)+\partial^i P + v^i \dot{P}&=  - (\rho + P) \partial^i \Phi -2 \Psi \partial^i P,
\end{align}
\end{subequations}
and after neglecting $P$, we find
\begin{subequations}
\begin{align}\label{nonrelenergyconservation}
    \dot \rho +\partial_i(v^i\rho)  &= -2\rho v^i\partial_i \Phi,\\ \label{nonrelmomentumconservation}
    \rho (\dot v^i + v^j \partial_j v^i)+ \partial^i P&=  - \rho\partial^i \Phi .
\end{align}
\end{subequations}

For the superfluid case, we need to take into account the metric dependence of $X$ in \eqref{Tmunu}, which renders it not quite the energy-momentum tensor of a perfect fluid. We have
\begin{align}
    X &= -(1-2\Phi)\dot{\psi}^2 + (1+2\Psi) \partial^i \psi \partial_i \psi \nonumber\\
    &\approx -m^2 -2m\left(\dot{\theta} -\frac{1}{2m}(\nabla \theta)^2 -m \Phi\right) +2\Psi (\nabla \theta)^2,
\end{align}
while the four-velocity reads
\begin{align}
    \nonumber \tilde{u}^\mu &= \frac{1}{\sqrt{-X}}((1-2\Phi)(m+ \dot{\theta}),-(1+2\Psi)\partial^i \theta)\\
    &\approx \left(1 - \Phi,-(1+2\Psi) \frac{\partial^i \theta}{m}\right),
\end{align}
and to leading order in the metric perturbations these are the same as in the flat case. The first modification on $X$ is the shift $Y \to Y - m\Phi$. Using $v^2 \approx (\nabla \theta)^2/m^2 \ll 1 $, we have
\begin{equation}
    X \approx -m^2 - 2m\left(\dot{\theta} -\frac{m}{2} v^2 - m\Phi\right). 
\end{equation}
Thus, to leading order, we might assume $X$ and $\tilde{u}^\mu$ to be the same as in the flat case when taking the Newtonian limit of the theory.

However, since the metric is not flat, the equation of motion for $\theta$ will now give
\begin{equation}
    \nabla_\mu(\mathcal{P}_Y \tilde{u}^\mu) = \partial_\mu (\mathcal{P}_Y \tilde{u}^\mu) + \Gamma^\mu_{\mu\rho} \mathcal{P}_Y \tilde{u}^\rho = 0,
\end{equation}
where the non-vanishing independent Christofell symbols for the metric \eqref{newtonianmetric} are
\begin{align}
    \nonumber \Gamma_{i0}^0 &= \partial_i \Phi,   &   \Gamma_{00}^j &= \partial^j \Phi,\\
    \Gamma_{kl}^j &= \eta_{kl}\partial^j \Psi - 2{\delta^j}_{(k} {\partial}_{l)} \Psi,
\end{align}
such that $\Gamma_{\mu 0}^\mu = 0$ and $\Gamma_{\mu l}^\mu = \partial_{l}(\Phi - 3 \Psi)$. So, collecting the leading order terms of $\tilde{u}^\mu$, we find
\begin{align}
    \nonumber \partial_\mu (\mathcal{P}_Y \tilde{u}^\mu ) + \mathcal{P}_Y\Gamma_{\mu\rho}^\mu \tilde{u}^\rho &\approx  \partial_0 \mathcal{P}_Y - \frac{1}{m}\partial_{i}(\mathcal{P}_Y \partial^i \theta)\\
    &\;\;\:- \mathcal{P}_Y\frac{1}{m}\partial^i \theta \partial_i (\Phi - 3\Psi) = 0,\label{nonreleq1}
\end{align}
which is just the statement of energy conservation \eqref{nonrelenergyconservation}, after setting $\Phi = \Psi$. Momentum conservation \eqref{nonrelmomentumconservation} implies
\begin{equation}\label{nonreleq2}
    m \mathcal{P}_Y\left(-\frac{\partial^i \dot \theta}{m}+ \frac{\partial^j \theta}{m^2}\partial_j \partial^i \theta\right) \approx - \partial^i \mathcal{P} - m \mathcal{P}_Y \partial^i \Phi .
\end{equation}
Note that, in equations \eqref{nonreleq1} and \eqref{nonreleq2}, $Y$ is taken to be the flat one. To leading order, the condition for equilibrium, i.e. $\partial^i \theta =0$ such that $Y = \dot \theta = \mu_{\text{n.r.}}$, is
\begin{equation}
    \partial^i \mathcal{P} = - m \mathcal{P}_Y \partial^i \Phi \implies \partial^i Y = -m \partial^i \Phi.
\end{equation}
For a spherical configuration, we can integrate the Poisson equation to find
\begin{align}
    \partial_r \Phi &= \frac{4\pi G M(r)}{r^2},\\
    \quad M(r) &= \int_0^r d\tilde{r}\; \tilde{r}^2 \rho(\tilde{r}) = \int_0^r d\tilde{r}\; \tilde{r}^2 m \mathcal{P}_Y(Y(\tilde{r})),
\end{align}
such that
\begin{equation}
    \nabla^2 Y = -m \nabla^2 \Phi = -4\pi G m \rho \implies \partial_r Y=  - \frac{4\pi G mM(r)}{r^2},
\end{equation}
which can be solved once an equation of state $P = \mathcal{P}(Y)$ is given.

In superfluid DM, one typically assumes that the baryons self-interact via phonon exchange. This plays a major role in explaining some correlations between the acceleration in galaxies and their baryonic mass, which can also be explained by MOND. A typical equation of state that can mimic the required MOND dynamics to fit some galactic rotation curves is \cite{Berezhiani:2015bqa}
\begin{equation}
    \mathcal{P}(Y) = \frac{2}{3}\Lambda (2m)^{3/2}Y \sqrt{|Y|},
\end{equation}
where $\Lambda$ is the energy scale associated to the baryon-photon coupling term. For a homogeneous BEC, 
\begin{equation}
    \mathcal{P} = \frac{2}{3}\Lambda (2m\mu)^{3/2} = \frac{\rho^3}{12\Lambda^2 m^6}. 
\end{equation}

Including gravity, we can estimate the size of a spherical condensed region. The Poisson equation above gives
\begin{equation}
    \partial_r (r^2 \partial_r \mu) = -4\pi Gm [m\Lambda (2m)^{3/2}]r^2 \mu^{1/2},
\end{equation}
which is essentially a Lane-Emden equation. The following numerical solution with the boundary conditions $\rho(0) = \rho_{\text{core}}$ and $\partial_r \rho(0) =0$ was found in \cite{Berezhiani:2015bqa}:
\begin{align}
    \rho_{\text{core}} &\approx 5.5\frac{3M_{\text{core}}}{4\pi R^3_{\text{core}}},\\
    \quad R_{\text{core}} &\approx \left(\frac{M_{\text{core}}}{10^{11}M_{\odot}}\right)^{1/5}\left(\frac{m}{\text{eV}}\right)^{-6/5}\left(\frac{\Lambda}{\text{meV}}\right)^{-2/5}45 \text{kpc}.
\end{align}
Thus, we can have the spherical condensate extending throughout realistic galactic cores for $\Lambda \sim \text{meV}$ and $m\sim \text{eV}$.

Let's now estimate the sound speed on these spherical condensates. We have
\begin{equation}
    c_s^2 = \frac{\partial P}{\partial \rho} \approx \frac{\rho^2}{4\Lambda^2 m^6},
\end{equation}
and evaluating at the core density with $\Lambda \sim \text{meV}$ and $m \sim \text{eV}$, we find
\begin{equation}
    c_s \sim 1.6 \times 10^3 \text{m}/\text{s} \sim 10^{-5}c. 
\end{equation}
We conclude that the typical sound speeds in DM superfluid condensates are nonrelativistic, as expected.

\subsection{High-temperature Limit}
\label{ssec:ht}

In this section, we shall consider temperatures higher or slightly higher than the critical value, $T \gtrsim T_{c}$, such that the superfluid component disappears ($\tilde{u}^{\mu} = 0$) and we are left with a perfect fluid. The dependence of $F$ on $X$ can be neglected, i.e. $F = F\left(b, y\right)$, and the energy-momentum tensor becomes
\begin{align}
    T_{\mu\nu} = \left(y F_{y} - b F_{b}\right) u_{\mu} u_{\nu} + \left(F - b F_{b}\right) g_{\mu\nu},\label{eq:e01}
\end{align}
while the conserved current $N^{\mu}$ reduces to
\begin{align}
    N^{\mu} = \frac{\partial F}{\partial\left(\partial_{\mu}\psi\right)} = F_{y} u^{\mu}.\label{eq:e14}
\end{align}

To obtain the thermodynamic variables, we first contract $T_{\mu \nu}$ with the projector $\left(g^{\mu \nu} + u^{\mu} u^{\nu}\right)$, which gives
\begin{align}
	P = \frac{1}{3} \left(g^{\mu\nu} + u^{\mu} u^{\nu}\right) T_{\mu\nu} = F - b F_{b}.\label{eq:e15}
\end{align}

The energy density is obtained by contracting $T^{\mu\nu}$ with the four velocity,
\begin{align}
	\rho = T_{\mu\nu} u^{\mu} u^{\nu} = y F_{y} - F,\label{eq:e16}
\end{align}
and the particle number density comes from the current $N^{\mu}$,
\begin{align}
	n = - N^{\mu} u_{\mu} = F_{y}.\label{eq:e17}
\end{align}
The remaining thermodynamic variables are the same as in the general case.

With these definitions, the energy-momentum tensor defined in \eqref{eq:e01} can be rewritten as
\begin{align}
    T^{\mu\nu} = \left(P + \rho\right) u^{\mu} u^{\nu} + g^{\mu\nu} P.\label{eq:e02}
\end{align}

Now, we want to recover the nonrelativistic limit for $T \gtrsim T_{c}$. Similarly to what was done for the low-temperature case, we write the comoving coordinates of the normal component as $\phi^{i} = \alpha x^{i} + \zeta^{i}$, where $x^{i}$ are the physical coordinates, $\partial_{i}\zeta^{i} \ll \alpha$, and $\alpha$ is a constant whose physical meaning will become evident in the following discussion. In terms of $x^{i}$ and $\zeta^{i}$, $J^{\mu}$ becomes
\begin{align}
    \nonumber J^{\mu} &= \frac{1}{6} \epsilon^{\mu\sigma\beta\gamma} \epsilon_{ijk} \left(\alpha \partial_{\sigma}x^{i} + \partial_{\sigma}\zeta^{i}\right) \left(\alpha \partial_{\beta}x^{j} + \partial_{\beta}\zeta^{j}\right) \left(\alpha \partial_{\gamma}x^{k} + \partial_{\gamma}\zeta^{k}\right)\\
    &= - \alpha^{3} g^{\mu 0} + \frac{1}{2} \alpha^{2} \epsilon^{\mu i j \gamma} \epsilon_{ijk} \partial_{\gamma}\zeta^{k} + \frac{1}{2} \alpha \epsilon^{\mu i \beta \gamma} \epsilon_{ijk} \partial_{\beta}\zeta^{j} \partial_{\gamma}\zeta^{k} + \mathcal{O}\left(\zeta^{3}\right),
\end{align}
and $b$ becomes
\begin{align}
    \nonumber b &= \sqrt{- J^{\mu} J_{\mu}} \approx \alpha^{3} \left[1 + \frac{\partial_{i}\zeta^{i}}{\alpha} - \frac{1}{2} \left(\frac{\dot{\zeta}^{i}}{\alpha}\right)^{2}\right]\\
    &\approx \alpha^{3} + \alpha^{2} \left[\partial_{i}\zeta^{i} - \frac{1}{2 \alpha} \left(\dot{\zeta}^{i}\right)^{2}\right] \equiv \alpha^{3} + \alpha^{2} Z,
\end{align}
where we defined the new variable
\begin{align}
	Z = \partial_{i}\zeta^{i} - \frac{\left(\dot{\zeta}^{i}\right)^{2}}{2 \alpha}.
\end{align}

We can rewrite the four-velocity of the normal component as $ u^{\mu} = - \dfrac{J^{\mu}}{b}$, so that
\begin{align}
    u^{0} &\approx 1 - \frac{1}{2} \left(\frac{\dot{\zeta}^{i}}{\alpha}\right)^{2} - \frac{1}{2} \left(\frac{\dot{\zeta}^{i}}{\alpha}\right)^{2} \frac{\partial_{j}\zeta^{j}}{\alpha},\\
    u^{i} &\approx \frac{\dot{\zeta}^{i}}{\alpha} - \frac{\partial_{j}\zeta^{i}}{\alpha} \frac{\dot{\zeta}^{j}}{\alpha} + \frac{1}{2} \frac{\dot{\zeta}^{i}}{\alpha} \left(\frac{\dot{\zeta}^{j}}{\alpha}\right)^{2} + \frac{1}{2} \frac{\dot{\zeta}^{i}}{\alpha} \left(\frac{\dot{\zeta}^{j}}{\alpha}\right)^{2} \frac{\partial_{k}\zeta^{k}}{\alpha} - \frac{1}{2} \frac{\partial_{j}\zeta^{i}}{\alpha} \frac{\dot{\zeta}^{j}}{\alpha} \left(\frac{\dot{\zeta}^{k}}{\alpha}\right)^{2},
\end{align}
up to first order in $\partial_{i}\zeta^{i}$. Now, we know that in the nonrelativistic limit the four-velocity should reduce to $u^{\mu} \approx \left(1, v^{i}\right)$, with $v^{i} \ll 1$; this can be achieved by taking $\dot{\zeta}^{i} \ll \alpha$, for which
\begin{align}
    u^{\mu} &\approx \left(1, \frac{\dot{\zeta}^{i}}{\alpha}\right) \equiv \left(1, v^{i}\right),
\end{align}
where $v^{i} = \dot{\zeta}^{i}/\alpha$. Similarly to \eqref{eq:e06}, we can rewrite $Z$ in terms of $v^{i}$:
\begin{align}
    Z = \partial_{i}\zeta^{i} - \frac{1}{2} \alpha v^{2}.
\end{align}
In this limit, $b \approx - \alpha^{3}$, and comparing it to equation \eqref{eq:e07}, we see that $\alpha$ and the entropy are related by $s = \alpha^{3}$.

Finally, $y$ is given by
\begin{align}
    \nonumber y &= u^{\mu} \partial_{\mu}\psi = u^{0} \partial_{0}\psi + u^{i} \partial_{i}\psi \approx m + \dot{\theta} + \frac{\dot{\zeta}^{i}}{\alpha} \partial_{i}\theta\\
    &= m + \dot{\theta} - m v^{i} \tilde{v}_{i}
    \approx m,
\end{align}
where we have once again used $\psi = m t + \theta$.

In the nonrelativistic limit, the pressure $P = F - b F_{b}$ is much smaller than the energy density $\rho = y F_{y} - F$; this can be achieved by taking $b F_{b} \ll y F_{y}$ and $F \ll y F_{y}$, so that
\begin{align}
	\rho \approx y F_{y} \approx m n,
\end{align}
as expected.

Hence, in this regime, $F\left(b, y\right)$ should be viewed as a function of $Z$, $F\left(b\left(Z\right), y\right) \equiv Q\left(Z, y\right)$, such that
\begin{align}
    F_{b} = \frac{d Z}{d b} Q_{Z} = \frac{Q_{Z}}{\alpha^{2}}, \quad F_{y} = Q_{y},
\end{align}
and the thermodynamic variables become
\begin{align}
	\rho &\approx m Q_{y} - Q \approx m Q_{y},	&	P &\approx Q - \alpha Q_{Z} \approx 0,\\
	n &\approx Q_{y},	&	N^{\mu} &\approx Q_{y} u^{\mu}.
\end{align}

The conservation of the energy-momentum tensor in flat spacetime leads to
\begin{align}
	\nonumber 0 &= \partial_{\mu} T^{\mu\nu} = \partial_{\mu} \left[\left(y F_{y} - b F_{b}\right) u^{\mu} u^{\nu}\right] + g^{\mu\nu} \partial_{\mu}\left(F - b F_{b}\right)\\
	&= \partial_{\mu} \left[\left(\rho + P\right) u^{\mu} u^{\nu}\right] + \partial^{\nu}P.
\end{align}
For $\nu = 0$, we have
\begin{align}
	\nonumber 0 &= \partial_{\mu} \left[\left(\rho + P\right) u^{\mu} u^{0}\right] + \partial^{0}P = \partial_{\mu} \left[\left(\rho + P\right) u^{\mu}\right] - \partial_{0}P\\
	\nonumber &= \partial_{0} \left[\left(\rho + P\right) u^{0}\right] + \partial_{i} \left[\left(\rho + P\right) v^{i}\right] - \partial_{0}P = \partial_{0}\left(\rho + P\right) + \partial_{i} \left[\left(\rho + P\right) v^{i}\right] - \partial_{0}P\\
	&= \partial_{0}\rho + \partial_{i} \left[\left(\rho + P\right) v^{i}\right] = \frac{\partial\rho}{\partial t} + \partial_{i} \left[\left(\rho + P\right) v^{i}\right],
\end{align}
and for $\nu = i$,
\begin{align}
	\nonumber 0 &= \partial_{\mu} \left[\left(\rho + P\right) u^{\mu} u^{i}\right] + \partial^{i}P = \partial_{0}\left[\left(\rho + P\right) u^{0} v^{i}\right] + \partial_{j}\left[\left(\rho + P\right) v^{i} v^{j}\right] + g^{ij} \partial_{j}P\\
	&= \partial_{0}\left[\left(\rho + P\right) v^{i}\right] + \partial_{j}\left[\left(\rho + P\right) v^{i} v^{j} + g^{ij} P\right].
\end{align}
Considering $P \ll \rho$, we obtain
\begin{align}
	\frac{\partial\rho}{\partial t} + \partial_{i} \left(\rho v^{i}\right) \equiv \frac{\partial\rho}{\partial t} + \partial_{i} j^{i} = 0,\\
	\frac{\partial}{\partial t}\left(\rho v^{i}\right) + \frac{\partial}{\partial x^{k}}\left(\rho v^{i} v^{k} + g^{ik} P\right) \equiv \frac{\partial j^{i}}{\partial t} + \frac{\partial \Pi^{ik}}{\partial x^{k}} = 0,
\end{align}
i.e. the continuity and momentum conservation equations.

To obtain the equation of entropy conservation, we replace the energy density $\rho$ in \eqref{eq:e02} by the thermodynamic identity $\rho = T s + \mu n - P$,
\begin{align}
	T^{\mu\nu} = \left(T s + \mu n\right) u^{\mu} u^{\nu} + g^{\mu\nu} P,
\end{align}
and requiring its conservation in flat spacetime, we find
\begin{align}
	\nonumber 0 &= \partial_{\mu}T^{\mu\nu} = \partial_{\mu}\left(T s u^{\mu} + \mu n u^{\mu}\right) u^{\nu} + \left(T s + \mu n\right) u^{\mu} \partial_{\mu}u^{\nu} + \partial^{\nu}P\\
	&= \left[\left(s \partial_{\mu}T + n \partial_{\mu}\mu\right) u^{\mu} + T \partial_{\mu}\left(s u^{\mu}\right) + \mu \partial_{\mu}\left(n u^{\mu}\right)\right] u^{\nu} + \left(T s + \mu n\right) u^{\mu} \partial_{\mu}u^{\nu} + \partial^{\nu}P,
\end{align}
and using the Gibbs-Duhem relation, $s \partial_{\mu}T + n \partial_{\mu}\mu = \partial_{\mu}P$, together with particle number conservation, $\partial_{\mu}N^{\mu} = \partial_{\mu}\left(n u^{\mu}\right) = 0$, we find
\begin{align}
	0 &= \left[u^{\mu} \partial_{\mu}P + T \partial_{\mu}\left(s u^{\mu}\right)\right] u^{\nu} + \left(T s + \mu n\right) u^{\mu} \partial_{\mu}u^{\nu} + \partial^{\nu}P.
\end{align}
Contracting this expression with $u_{\nu}$ yields
\begin{align}
	\nonumber 0 &= \left[u^{\mu} \partial_{\mu}P + T \partial_{\mu}\left(s u^{\mu}\right)\right] u^{\nu} u_{\nu} + \left(T s + \mu n\right) u_{\nu} u^{\mu} \partial_{\mu}u^{\nu} + u_{\nu} \partial^{\nu}P\\
	\nonumber &= - u^{\mu} \partial_{\mu}P - T \partial_{\mu}\left(s u^{\mu}\right) + \left(T s + \mu n\right) u_{\nu} u^{\mu} \partial_{\mu}u^{\nu} + u_{\nu} \partial^{\nu}P\\
	\nonumber &= - T \partial_{\mu}\left(s u^{\mu}\right)\\
	\therefore \;&\partial_{\mu}\left(s u^{\mu}\right) = 0,
\end{align}
where we have used $u^{\nu} u_{\nu} = - 1 \;\Longrightarrow \; u_{\nu} \partial_{\mu}u^{\nu} = 0$.

\section{Recovering the nonrelativistic dynamics for \ensuremath{T < T\texorpdfstring{\textsubscript{c}}{c}}}
\label{sec:map}

In order to take the nonrelativistic limit of the theory at any temperature below $T_{c}$, we start by assuming that, in this limit, the Lagrangian $F$ can be decomposed as
\begin{align}
    F\left(X, y, b\right) \to F_{S}\left(X\right) + F_{N}\left(y, b\right),
\end{align}
where the subscripts $S$ and $N$ denote the superfluid and normal parts of $F$, respectively. This ansatz is motivated by the fact that, to leading order, $X$ is proportional to $m^2$, which is just a parameter. As we shall see, this ansatz yields a nonrelativistic limit that matches the two-fluid model. With this assumption, the derivatives of $F$ with respect to the variables $y$, $X$ and $b$ become, respectively,
\begin{align}
    F_{y} &= \left(F_{N}\right)_{y},   &   F_{X} &= \left(F_{S}\right)_{X},    &   F_{b} &= \left(F_{N}\right)_{b},
\end{align}
where we made use of the fact that the normal fluid depends on the variables $y$ and $b$, while the superfluid only depends on $X$.

Since the energy-momentum tensor is linear in the Lagrangian and its derivatives, all the hydro- and thermodynamic variables will also decompose linearly into superfluid and normal contributions. The thermodynamic variables can be written as
\begin{align}
	\nonumber N^{\mu} &= - 2 \sqrt{- X} F_{X}\left(X, y, b\right) \tilde{u}^{\mu} + F_{y}\left(X, y, b\right) u^{\mu}\\
	&= - 2 \sqrt{- X} \left(F_{S}\right)_{X} \tilde{u}^{\mu} + \left(F_{N}\right)_{y} u^{\mu},\\
	\nonumber P &= F\left(X, y, b\right) - b F_{b}\left(X, y, b\right)\\
	&= F_{N} + F_{S} - b \left(F_{N}\right)_{b},\\
	\nonumber \rho &= y F_{y}\left(X, y, b\right) - 2 y^{2} F_{X}\left(X, y, b\right) - F\left(X, y, b\right)\\
	&= y \left(F_{N}\right)_{y} - 2 y^{2} \left(F_{S}\right)_{X} - F_{N} - F_{S},\\
	\nonumber n &= F_{y}\left(X, y, b\right) - 2 y F_{X}\left(X, y, b\right)\\
	&= \left(F_{N}\right)_{y} - 2 y \left(F_{S}\right)_{X}
\end{align}
for the system between the $T = 0$ and $T = T_{c}$. From these equations, we can identify the superfluid and normal variables as follows:
\begin{align}
    N^{\mu}_{N} &= \left(F_{N}\right)_{y} u^{\mu},  &   N^{\mu}_{S} &= - 2 \sqrt{- X} \left(F_{S}\right)_{X} \tilde{u}^{\mu},\\
    P_{N} &= F_{N} - b \left(F_{N}\right)_{b},  &   P_{S} &= F_{S},\\
    \rho_{N} &= y \left(F_{N}\right)_{y} - F_{N},   &   \rho_{S} &= - 2 y^{2} \left(F_{S}\right)_{X} - F_{S},\\
    n_{N} &= \left(F_{N}\right)_{y},    &   n_{S} &= - 2 y \left(F_{S}\right)_{X}.
\end{align}
As expected, the normal component's thermodynamic variables are identical to those computed in Section~\ref{ssec:ht} in the high-temperature limit, \eqref{eq:e14}, \eqref{eq:e15}, \eqref{eq:e16}, \eqref{eq:e17}; the remaining variables, however, do not match completely the results from section~\ref{ssec:lt}, due to the fact that, at $T = 0$, the dependence of $F$ on $y$ and $b$ disappear. As consistency check, we can recover the low-temperature variables \eqref{eq:e13} by replacing $y^{2} \longrightarrow - X$, so that
\begin{align}
    \rho_{S} &= 2 X \left(F_{S}\right)_{X} - F_{S}, &   n_{S} &= - 2 \sqrt{- X} \left(F_{S}\right)_{X}.
\end{align}

As was done in the previous sections, in the nonrelativistic limit, we can write $X$ and $b$ in terms of $Y$ and $Z$,
\begin{align}
	X &= - m^{2} - 2 m Y,  &   b &= \alpha^{3} + \alpha^{2} Z,
\end{align}
and since $X = - y^{2} + \xi^{2}$ and $\xi^{2} = y^{2} \left(v^{i} - \tilde{v}^{i}\right)^{2} \equiv y^{2} \left(w^{i}\right)^{2}$, we can write $y$ as
\begin{align}
	\nonumber y^{2} &= - \frac{X}{1 - w^{2}} = \frac{m^{2} + 2 m Y}{1 - w^{2}} = m^{2} \left(\frac{1 + 2 Y/m}{1 - w^{2}}\right)\\
	y &\approx Y + m \left(1 + \frac{w^{2}}{2}\right).
\end{align}
Thus, we can write the derivatives of $F$ as
\begin{align}
    \nonumber F_{X} &= \frac{\partial F}{\partial Y} \frac{dY}{dX} = - \frac{F_{Y}}{2 m},    &   F_{b} &= \frac{\partial F}{\partial Z} \frac{dZ}{db} = \frac{F_{Z}}{\alpha^{2}},\\
    F_{y} &= \frac{\partial F}{\partial Y} \frac{dY}{dy} + \frac{\partial F}{\partial w} \frac{dw}{dy} = F_{Y} + \frac{F_{w}}{m w},
\end{align}
so that
\begin{align}
    N^{\mu}_{N} &\approx \left[\left(F_{N}\right)_{Y} + \frac{\left(F_{N}\right)_{w}}{m w}\right] u^{\mu},   &   N^{\mu}_{S} &\approx \left(1 + \frac{Y}{m}\right) \left(F_{S}\right)_{Y} \tilde{u}^{\mu},\\
    P_{N} &\approx F_{N} - \alpha \left(1 + \frac{Z}{\alpha}\right) \left(F_{N}\right)_{Z},    &   P_{S} &= F_{S},\\
    \rho_{N} &\approx \left(1 + \frac{w^{2}}{2} + \frac{Y}{m}\right) \left[m \left(F_{N}\right)_{Y} + \frac{\left(F_{N}\right)_{w}}{w}\right] - F_{N},    &    \rho_{S} &\approx m \left(1 + w^{2} + \dfrac{2 Y}{m}\right) \left(F_{S}\right)_{Y} - F_{S},\\
    n_{N} &\approx \left(F_{N}\right)_{Y} + \frac{\left(F_{N}\right)_{w}}{m w},    &   n_{S} &\approx \left(1 + \dfrac{w^{2}}{2} + \dfrac{Y}{m}\right) \left(F_{S}\right)_{Y}.
\end{align}
Taking $Y/m \ll 1$, $Z/\alpha \ll 1$ and $w \ll 1$, as well as $P \ll \rho$, we obtain
\begin{align}
    N^{\mu}_{N} &\approx \left[\left(F_{N}\right)_{Y} + \frac{\left(F_{N}\right)_{w}}{m w}\right] u^{\mu},  &   N^{\mu}_{S} &\approx \left(F_{S}\right)_{Y} \tilde{u}^{\mu},\\
     P_{N} &\approx 0,  &   P_{S} &\approx 0,\\
    \rho_{S} &\approx m \left(F_{S}\right)_{Y},   &   \rho_{N} &\approx m \left(F_{N}\right)_{Y} + \frac{\left(F_{N}\right)_{w}}{w},\\
    n_{S} &\approx \left(F_{S}\right)_{Y},  &   n_{N} &\approx \left(F_{N}\right)_{Y} + \frac{\left(F_{N}\right)_{w}}{m w},
\end{align}
and we can see that $\rho_{S} \approx m n_{S}$ and $\rho_{N} \approx m n_{N}$.
	
With those expressions in hand, we can regroup the terms of $T^{\mu\nu}$ as
\begin{align}
    \nonumber T^{\mu\nu} &= 2 X \left(F_{S}\right)_{X} \tilde{u}^{\mu} \tilde{u}^{\nu} + \left[y \left(F_{N}\right)_{y} - b \left(F_{N}\right)_{b}\right] u^{\mu} u^{\nu} + \left[F_{N} + F_{S} - b \left(F_{N}\right)_{b}\right] g^{\mu\nu}\\
    \nonumber &= \left[\left(\rho_{S} + P_{S}\right) \tilde{u}^{\mu} \tilde{u}^{\nu} + P_{S} g^{\mu\nu}\right] + \left[\left(\rho_{N} + P_{N}\right) u^{\mu} u^{\nu} + P_{N} g^{\mu\nu}\right]\\
    &\equiv T^{\mu\nu}_{S} + T^{\mu\nu}_{N},
\end{align}
where the linearity of the energy-momentum tensor in $F$ and its derivatives made it so it can be easily decomposed into normal and superfluid parts.

We now consider the nonrelativistic dynamics of our theory, in which all the particles velocities are small compared to that of light, and try to recover the fluid equations \eqref{eq:e04}, \eqref{eq:e05}, \eqref{eq:e08}, and \eqref{eq:e10}. Those equations are written in terms of the mass density, $\rho_{m}$, the velocities of the normal and superfluid components, $\vec{v}_{n}$ and $\vec{v}_{s}$, and the energy density $\epsilon$. The latter does not include the contribution from the rest-mass energy. Before delving into the dynamical equations, let us then discuss how $\epsilon$ is recovered in the nonrelativistic limit. In our notation, $\rho$ stands for the relativistic energy density, which includes contributions of the rest-mass and kinetic energies. In order to recover the fluid equations, we need to split $\rho$ into a mass density term, $\rho_{m}$, and an energy density $\epsilon$.

As an example, consider a dust-like gas of massive particles, such that
\begin{align}
    \rho = \gamma m_{0} n,
\end{align}
where $\gamma = \left(1 - v^{2}\right)^{-1/2}$ is the usual Lorentz gamma factor, $m_{0}$ denotes the rest mass of the particle, and $n$ is the particle number density.  Starting from the relativistic momentum density,
\begin{align}
    \vec{j} = \gamma m_{0} n \vec{v} = \rho \vec{v},
\end{align}
we can obtain an expression for $v^{2}$,
\begin{align}
    v^{2} = \frac{j^{2}}{m_{0}^{2} n^{2} + j^{2}},
\end{align}
which can then be replaced in the expression for $\gamma$, leading to
\begin{align}
    \gamma &= \frac{1}{\sqrt{1 - \dfrac{j^{2}}{m_{0}^{2} n^{2} + j^{2}}}} = \sqrt{1 + \left(\frac{j}{m_{0} n}\right)^{2}}.
\end{align}
Plugging $\gamma$ back into the definition of $\rho$, we obtain
\begin{align}
    \rho = m_{0} n \sqrt{1 + \left(\frac{j}{m_{0} n}\right)^{2}}.
\end{align}
In the nonrelativistic limit, the four-velocities of the normal and superfluid become, respectively, $u^{\mu} \sim \left(1, v^{i}\right)$ and $\tilde{u}^{\mu} \sim \left(1, \tilde{v}^{i}\right)$, with $v$, $\tilde{v} \ll 1$, and we can expand our quantities in powers of $v$ and $\tilde{v}$. The Lorentz factor can be expanded as
\begin{align}
	\gamma \sim 1 + \frac{1}{2} v^{2} + \mathcal{O}\left(v^{4}\right),
\end{align} 
so that, at leading order, we have
\begin{align}
    \rho \approx m_{0} n + \frac{1}{2} \frac{j^{2}}{m_{0} n} \approx m_{0} n + \frac{m_{0} n v^{2}}{2} = \rho_{m} + \epsilon.
\end{align}
As can be seen from this expression, the energy density $\epsilon$ is of the order $v^{2}$. Moreover, the pressure is of the same order of magnitude as $\epsilon$. This can be demonstrated using the case of an ideal gas: the internal energy density is given by $\epsilon = n c_{v} k_{B} T$, where $c_{v} \sim 1$ is the specific heat capacity at constant volume, while the pressure is $P = n k_{B} T$. From this example, we see that $\epsilon = \rho - \rho_m$, and this will also be true in general. For more complex fluids, $\epsilon$ will also include contributions from the internal energy of its constituents.

Now that we know how to identify $\rho_{m}$ and $\epsilon$ in our expressions, Landau's two-fluid equations can be recovered.

As showed, the particle number four-current can be decomposed into a normal and a superfluid parts, $N^{\mu} = N_{N}^{\mu} + N_{S}^{\mu}$, so that its conservation leads to
\begin{align}
    \nonumber 0 &= \partial_{\mu}N^{\mu} = \partial_{\mu}\left(N_{N}^{\mu} + N_{S}^{\mu}\right) = \partial_{\mu}\left(n_{N} u^{\mu} + n_{S} \tilde{u}^{\mu}\right)\\
    \nonumber &= \partial_{0}\left(n_{N} + n_{S}\right) + \partial_{i}\left(n_{N} v^{i} + n_{S} \tilde{v}^{i}\right) = \partial_{0}n + \partial_{i}\left(N_{N}^{i} + N_{S}^{i}\right)\\
    &= \partial_{0}n + \partial_{i}N^{i},
\end{align}
where $n = n_{N} + n_{S}$ is the total particle number density. Multiplying the expression by $m$, and using $\rho_{m} = m n$, we recover the continuity equation,
\begin{align}
    \frac{\partial\rho_{m}}{\partial t} + \vec{\nabla} \cdot \vec{j} = 0.
\end{align}

To obtain the equation of momentum conservation, we start from the conservation of the energy-momentum tensor, $\partial_{\mu} T^{\mu\nu} = 0$, and take $\nu = j$:
\begin{align}
    \nonumber 0 &= \partial_{\mu} T^{\mu j} = \partial_{\mu}\left(T^{\mu j}_{S} + T^{\mu j}_{N}\right) = \partial_{\mu}\left[\left(\rho_{S} + P_{S}\right) \tilde{u}^{\mu} \tilde{u}^{j} + P_{S} g^{\mu j} + \left(\rho_{N} + P_{N}\right) u^{\mu} u^{j} + P_{N} g^{\mu j}\right]\\
    \nonumber &\equiv \partial_{\mu}\left[\left(\rho_{S} + P_{S}\right) \tilde{u}^{\mu} \tilde{v}^{j} + \left(\rho_{N} + P_{N}\right) u^{\mu} v^{j} + P g^{\mu j}\right]\\
    \nonumber &= \partial_{0}\left[\left(\rho_{S} + P_{S}\right) \tilde{v}^{j} + \left(\rho_{N} + P_{N}\right) v^{j}\right] +
    \partial_{i}\left[\left(\rho_{S} + P_{S}\right) \tilde{v}^{i} \tilde{v}^{j} + \left(\rho_{N} + P_{N}\right) v^{i} v^{j} + P g^{i j}\right],
\end{align}
where $P = P_{S} + P_{N}$ is the total pressure. Replacing the relativistic energy density by $\rho = \rho_{m} + \epsilon$, we obtain
\begin{align}
    \nonumber 0 &= \partial_{i}\left[\left(\rho_{S,m} + \epsilon_{S} + P_{S}\right) \tilde{v}^{i} \tilde{v}^{j} + \left(\rho_{N,m} + \epsilon_{N} + P_{N}\right) v^{i} v^{j} + P g^{i j}\right]\\
    \nonumber &\;\;\:+ \partial_{0}\left[\left(\rho_{S,m} + \epsilon_{S} + P_{S}\right) \tilde{v}^{j} + \left(\rho_{N,m} + \epsilon_{N} + P_{N}\right) v^{j}\right]\\
    &= \partial_{0}\left(\rho_{S,m} \tilde{v}^{j} + \rho_{N,m} v^{j}\right) + \partial_{i}\left(\rho_{S,m} \tilde{v}^{i} \tilde{v}^{j} + \rho_{N,m} v^{i} v^{j} + P g^{i j}\right) + \mathcal{O}\left(v^{3}\right).
\end{align}
Keeping only terms up to second order in $v$, we recover the momentum conservation equation,
\begin{align}
    \nonumber 0 &= \partial_{0}\left(\rho_{S,m} \tilde{v}^{k} + \rho_{N,m} v^{k}\right) + \partial_{i}\left(\rho_{S,m} \tilde{v}^{i} \tilde{v}^{k} + \rho_{N,m} v^{i} v^{k} + P g^{i k}\right)\\
    &= \frac{\partial j^{k}}{\partial t} + \frac{\partial \Pi^{ik}}{\partial x^{i}},
\end{align}
where $\vec{j}$ is the mass flux density, defined in equation \eqref{eq:e09}, and $\Pi^{ij}$ is the momentum flux density tensor, defined in equation \eqref{eq:e11}.

Taking, instead, $\nu = 0$ in $\partial_{\mu} T^{\mu\nu} = 0$, we find
\begin{align}
    \nonumber 0 &= \partial_{\mu} T^{\mu 0} = \partial_{\mu}\left(T_{s}^{\mu 0} + T_{N}^{\mu 0}\right) = \partial_{\mu}\left[\left(\rho_{S} + P_{S}\right) \tilde{u}^{\mu} + \left(\rho_{N} + P_{N}\right) u^{\mu} + P g^{\mu 0}\right]\\
    \nonumber &= \partial_{0}\left(\rho_{S} + P_{S} + \rho_{N} + P_{N} - P\right) +
    \partial_{i}\left[\left(\rho_{S} + P_{S}\right) \tilde{v}^{i} + \left(\rho_{N} + P_{N}\right) v^{i}\right]\\
    \nonumber &= \partial_{0}\rho +
    \partial_{i}\left[\left(\rho_{S} + P_{S}\right) \tilde{v}^{i} + \left(\rho_{N} + P_{N}\right) v^{i}\right]\\
    &= \partial_{0}\left(\rho_{m} + \epsilon\right) +
    \partial_{i}\left[\left(\rho_{S,m} + \epsilon_{S} + P_{S}\right) \tilde{v}^{i} + \left(\rho_{N,m} + \epsilon_{N} + P_{N}\right) v^{i}\right],
\end{align}
and the leading order contribution gives
\begin{align}
    \nonumber 0 &= \partial_{0}\rho_{m} +
    \partial_{i}\left(\rho_{S,m} \tilde{v}^{i} + \rho_{N,m} v^{i}\right) = \frac{\partial \rho_{m}}{\partial t} + \vec{\nabla} \cdot \vec{j},
\end{align}
i.e. we recover the continuity equation.

The next order contribution gives
\begin{align}
    \nonumber 0 &= \partial_{0}\epsilon +
    \partial_{i}\left[\left(\epsilon_{S} + P_{S}\right) \tilde{v}^{i} + \left(\epsilon_{N} + P_{N}\right) v^{i}\right]\\
    &= \frac{\partial\epsilon}{\partial t} +
    \partial_{i}\left[Q_{S}^{i} + Q_{N}^{i}\right] = \frac{\partial\epsilon}{\partial t} +
    \vec{\nabla} \cdot \vec{Q},
\end{align}
which is the energy conservation equation, and $\vec{Q}$ the energy flux density.

Lastly, the equation of conservation of entropy is an immediate consequence of the conservation of the four-current $J^{\mu}$, which we earlier identified with the entropy four-current:
\begin{align}
    0 &= \partial_{\mu}J^{\mu} = \partial_{\mu}\left(s u^{\mu}\right) = \frac{\partial s}{\partial t} + \vec{\nabla} \cdot \left(s \vec{v}\right).
\end{align}

\section{Conclusion} \label{sec:conc}

In this paper, we revisited two distinct frameworks for describing superfluidity: Landau's phenomenological two-fluid model and a fully relativistic effective field theory. We began by reviewing the two-fluid model, outlining the fundamental equations governing superfluid dynamics, followed by an overview of the key equations and variables from the effective field theory proposed in \cite{Nicolis:2011cs}. Our objective was to recover Landau's equations governing superfluid dynamics from the nonrelativistic limit of the effective field theory, thereby linking the general relativistic description with the nonrelativistic model commonly employed in cosmology.

We examined the nonrelativistic limit of the field theory, first in the low-temperature regime, followed by the high-temperature case. For the low-temperature limit, we also analyzed the behaviour of a self-gravitating superfluid. Next, we showed how to recover the nonrelativistic limit of the full theory, starting from the assumption that the Lagrangian can be decomposed into a superfluid and normal contributions in this limit. From there, we demonstrated that, due to the energy-momentum tensor being linear in the Lagrangian and its derivatives, all thermodynamic variables can be linearly decomposed into contributions from the superfluid and normal components. This allowed us to recover both the low- and high-temperature limit variables from the full theory ones. Finally, we showed that, given our chosen ansatz, Landau's equations are straightforwardly recovered.

\section*{Acknowledgement}

We thank Robert Brandenberger for many discussions and support during the execution of this work. HB also thanks Jorge Noronha for discussions. The research at McGill is supported in part by NSERC and the Canada Research Chair program.

\bibliographystyle{bibstyle} 
\bibliography{references.bib}

\providecommand{\href}[2]{#2}\begingroup\raggedright\begin{thebibliography}{10}

\bibitem{Lahav:2024npe}
O.~Lahav and A.~R. Liddle, \emph{{The Cosmological Parameters (2023)}},  \href{https://arxiv.org/abs/2403.15526}{{\ttfamily 2403.15526}}.

\bibitem{refId0}
{Planck Collaboration}, {Aghanim, N.}, {Akrami, Y.}, {Ashdown, M.}, {Aumont, J.}, {Baccigalupi, C.} et~al., \emph{Planck 2018 results - vi. cosmological parameters}, \href{https://doi.org/10.1051/0004-6361/201833910}{\emph{A\&A} {\bfseries 641} (2020) A6}.

\bibitem{hoekstra2004properties}
H.~Hoekstra, H.~K. Yee and M.~D. Gladders, \emph{Properties of galaxy dark matter halos from weak lensing}, {\emph{The Astrophysical Journal} {\bfseries 606} (2004) 67}.

\bibitem{bennett2013nine}
C.~L. Bennett, D.~Larson, J.~L. Weiland, N.~Jarosik, G.~Hinshaw, N.~Odegard et~al., \emph{Nine-year wilkinson microwave anisotropy probe (wmap) observations: final maps and results}, {\emph{The Astrophysical Journal Supplement Series} {\bfseries 208} (2013) 20}.

\bibitem{zwicky1933redshift}
F.~Zwicky et~al., \emph{The redshift of extragalactic nebulae}, {\emph{Helv. Phys. Acta} {\bfseries 6} (1933) 138}.

\bibitem{1936ApJ....83...23S}
S.~{Smith}, \emph{{The Mass of the Virgo Cluster}}, \href{https://doi.org/10.1086/143697}{\emph{The Astrophysical Journal} {\bfseries 83} (Jan., 1936) 23}.

\bibitem{1970ApJ...159..379R}
V.~C. {Rubin} and J.~{Ford}, W.~Kent, \emph{{Rotation of the Andromeda Nebula from a Spectroscopic Survey of Emission Regions}}, \href{https://doi.org/10.1086/150317}{\emph{The Astrophysical Journal} {\bfseries 159} (Feb., 1970) 379}.

\bibitem{1970ApJ...160..811F}
K.~C. {Freeman}, \emph{{On the Disks of Spiral and S0 Galaxies}}, \href{https://doi.org/10.1086/150474}{\emph{The Astrophysical Journal} {\bfseries 160} (June, 1970) 811}.

\bibitem{1973ApJ...186..467O}
J.~P. {Ostriker} and P.~J.~E. {Peebles}, \emph{{A Numerical Study of the Stability of Flattened Galaxies: or, can Cold Galaxies Survive?}}, \href{https://doi.org/10.1086/152513}{\emph{The Astrophysical Journal} {\bfseries 186} (Dec., 1973) 467--480}.

\bibitem{Anderson_2014}
L.~Anderson, E.~Aubourg, S.~Bailey, F.~Beutler, V.~Bhardwaj, M.~Blanton et~al., \emph{The clustering of galaxies in the sdss-iii baryon oscillation spectroscopic survey: baryon acoustic oscillations in the data releases 10 and 11 galaxy samples}, \href{https://doi.org/10.1093/mnras/stu523}{\emph{Monthly Notices of the Royal Astronomical Society} {\bfseries 441} (Apr., 2014) 24–62}.

\bibitem{Tegmark_2004}
M.~Tegmark, M.~R. Blanton, M.~A. Strauss, F.~Hoyle, D.~Schlegel, R.~Scoccimarro et~al., \emph{The three‐dimensional power spectrum of galaxies from the sloan digital sky survey}, \href{https://doi.org/10.1086/382125}{\emph{The Astrophysical Journal} {\bfseries 606} (May, 2004) 702–740}.

\bibitem{Bullock:2017xww}
J.~S. Bullock and M.~Boylan-Kolchin, \emph{{Small-Scale Challenges to the $\Lambda$CDM Paradigm}}, \href{https://doi.org/10.1146/annurev-astro-091916-055313}{\emph{Ann. Rev. Astron. Astrophys.} {\bfseries 55} (2017) 343--387}, [\href{https://arxiv.org/abs/1707.04256}{{\ttfamily 1707.04256}}].

\bibitem{weinberg2015cold}
D.~H. Weinberg, J.~S. Bullock, F.~Governato, R.~Kuzio~de Naray and A.~H. Peter, \emph{Cold dark matter: controversies on small scales}, {\emph{Proceedings of the National Academy of Sciences} {\bfseries 112} (2015) 12249--12255}.

\bibitem{751a}
D.~Spergel and P.~Steinhardt, \emph{Observational evidence for self-interacting cold dark matter}, \href{https://doi.org/10.1103/PhysRevLett.84.3760}{\emph{Physical review letters} {\bfseries 84} (2000) 3760--3763}.

\bibitem{Colin_2000}
P.~Colin, V.~Avila‐Reese and O.~Valenzuela, \emph{Substructure and halo density profiles in a warm dark matter cosmology}, \href{https://doi.org/10.1086/317057}{\emph{The Astrophysical Journal} {\bfseries 542} (Oct., 2000) 622–630}.

\bibitem{Berezhiani:2015bqa}
L.~Berezhiani and J.~Khoury, \emph{{Theory of dark matter superfluidity}}, \href{https://doi.org/10.1103/PhysRevD.92.103510}{\emph{Phys. Rev. D} {\bfseries 92} (2015) 103510}, [\href{https://arxiv.org/abs/1507.01019}{{\ttfamily 1507.01019}}].

\bibitem{Berezhiani:2015pia}
L.~Berezhiani and J.~Khoury, \emph{{Dark Matter Superfluidity and Galactic Dynamics}}, \href{https://doi.org/10.1016/j.physletb.2015.12.054}{\emph{Phys. Lett. B} {\bfseries 753} (2016) 639--643}, [\href{https://arxiv.org/abs/1506.07877}{{\ttfamily 1506.07877}}].

\bibitem{2015arXiv150703013K}
J.~{Khoury}, \emph{{A Dark Matter Superfluid}}, \href{https://doi.org/10.48550/arXiv.1507.03013}{\emph{arXiv e-prints} (July, 2015) arXiv:1507.03013}, [\href{https://arxiv.org/abs/1507.03013}{{\ttfamily 1507.03013}}].

\bibitem{Sin:1992bg}
S.-J. Sin, \emph{{Late time cosmological phase transition and galactic halo as Bose liquid}}, \href{https://doi.org/10.1103/PhysRevD.50.3650}{\emph{Phys. Rev. D} {\bfseries 50} (1994) 3650--3654}, [\href{https://arxiv.org/abs/hep-ph/9205208}{{\ttfamily hep-ph/9205208}}].

\bibitem{Goodman:2000tg}
J.~Goodman, \emph{{Repulsive dark matter}}, \href{https://doi.org/10.1016/S1384-1076(00)00015-4}{\emph{New Astron.} {\bfseries 5} (2000) 103}, [\href{https://arxiv.org/abs/astro-ph/0003018}{{\ttfamily astro-ph/0003018}}].

\bibitem{Peebles:2000yy}
P.~J.~E. Peebles, \emph{{Fluid dark matter}}, \href{https://doi.org/10.1086/312677}{\emph{Astrophys. J. Lett.} {\bfseries 534} (2000) L127}, [\href{https://arxiv.org/abs/astro-ph/0002495}{{\ttfamily astro-ph/0002495}}].

\bibitem{Hu:2000ke}
W.~Hu, R.~Barkana and A.~Gruzinov, \emph{{Cold and fuzzy dark matter}}, \href{https://doi.org/10.1103/PhysRevLett.85.1158}{\emph{Phys. Rev. Lett.} {\bfseries 85} (2000) 1158--1161}, [\href{https://arxiv.org/abs/astro-ph/0003365}{{\ttfamily astro-ph/0003365}}].

\bibitem{Silverman:2002qx}
M.~P. Silverman and R.~L. Mallett, \emph{{Dark matter as a cosmic Bose-Einstein condensate and possible superfluid}}, \href{https://doi.org/10.1023/A:1015934027224}{\emph{Gen. Rel. Grav.} {\bfseries 34} (2002) 633--649}.

\bibitem{Boehmer:2007um}
C.~G. Boehmer and T.~Harko, \emph{{Can dark matter be a Bose-Einstein condensate?}}, \href{https://doi.org/10.1088/1475-7516/2007/06/025}{\emph{JCAP} {\bfseries 06} (2007) 025}, [\href{https://arxiv.org/abs/0705.4158}{{\ttfamily 0705.4158}}].

\bibitem{Chavanis:2011zi}
P.-H. Chavanis, \emph{{Mass-radius relation of Newtonian self-gravitating Bose-Einstein condensates with short-range interactions: I. Analytical results}}, \href{https://doi.org/10.1103/PhysRevD.84.043531}{\emph{Phys. Rev. D} {\bfseries 84} (2011) 043531}, [\href{https://arxiv.org/abs/1103.2050}{{\ttfamily 1103.2050}}].

\bibitem{Bettoni:2013zma}
D.~Bettoni, M.~Colombo and S.~Liberati, \emph{{Dark matter as a Bose-Einstein Condensate: the relativistic non-minimally coupled case}}, \href{https://doi.org/10.1088/1475-7516/2014/02/004}{\emph{JCAP} {\bfseries 02} (2014) 004}, [\href{https://arxiv.org/abs/1310.3753}{{\ttfamily 1310.3753}}].

\bibitem{Guth:2014hsa}
A.~H. Guth, M.~P. Hertzberg and C.~Prescod-Weinstein, \emph{{Do Dark Matter Axions Form a Condensate with Long-Range Correlation?}}, \href{https://doi.org/10.1103/PhysRevD.92.103513}{\emph{Phys. Rev. D} {\bfseries 92} (2015) 103513}, [\href{https://arxiv.org/abs/1412.5930}{{\ttfamily 1412.5930}}].

\bibitem{Hui:2016ltb}
L.~Hui, J.~P. Ostriker, S.~Tremaine and E.~Witten, \emph{{Ultralight scalars as cosmological dark matter}}, \href{https://doi.org/10.1103/PhysRevD.95.043541}{\emph{Phys. Rev. D} {\bfseries 95} (2017) 043541}, [\href{https://arxiv.org/abs/1610.08297}{{\ttfamily 1610.08297}}].

\bibitem{Ferreira:2018wup}
E.~G.~M. Ferreira, G.~Franzmann, J.~Khoury and R.~Brandenberger, \emph{{Unified Superfluid Dark Sector}}, \href{https://doi.org/10.1088/1475-7516/2019/08/027}{\emph{JCAP} {\bfseries 08} (2019) 027}, [\href{https://arxiv.org/abs/1810.09474}{{\ttfamily 1810.09474}}].

\bibitem{Das:2014agf}
S.~Das and R.~K. Bhaduri, \emph{{Dark matter and dark energy from a Bose\textendash{}Einstein condensate}}, \href{https://doi.org/10.1088/0264-9381/32/10/105003}{\emph{Class. Quant. Grav.} {\bfseries 32} (2015) 105003}, [\href{https://arxiv.org/abs/1411.0753}{{\ttfamily 1411.0753}}].

\bibitem{Das:2018udn}
S.~Das and R.~K. Bhaduri, \emph{{Bose-Einstein condensate in cosmology}},  \href{https://arxiv.org/abs/1808.10505}{{\ttfamily 1808.10505}}.

\bibitem{Das:2022mgr}
S.~Das and S.~Sur, \emph{{A Unified Cosmological Dark Sector from a Bose-Einstein Condensate}},  \href{https://arxiv.org/abs/2203.16402}{{\ttfamily 2203.16402}}.

\bibitem{Ferreira:2020fam}
E.~G.~M. Ferreira, \emph{{Ultra-light dark matter}}, \href{https://doi.org/10.1007/s00159-021-00135-6}{\emph{Astron. Astrophys. Rev.} {\bfseries 29} (2021) 7}, [\href{https://arxiv.org/abs/2005.03254}{{\ttfamily 2005.03254}}].

\bibitem{Hui:2021tkt}
L.~Hui, \emph{{Wave Dark Matter}}, \href{https://doi.org/10.1146/annurev-astro-120920-010024}{\emph{Ann. Rev. Astron. Astrophys.} {\bfseries 59} (2021) 247--289}, [\href{https://arxiv.org/abs/2101.11735}{{\ttfamily 2101.11735}}].

\bibitem{Khoury:2021tvy}
J.~Khoury, \emph{{Dark Matter Superfluidity}}, \href{https://doi.org/10.21468/SciPostPhysLectNotes.42}{\emph{SciPost Phys. Lect. Notes} {\bfseries 42} (2022) 1}, [\href{https://arxiv.org/abs/2109.10928}{{\ttfamily 2109.10928}}].

\bibitem{Berezhiani:2017tth}
L.~Berezhiani, B.~Famaey and J.~Khoury, \emph{{Phenomenological consequences of superfluid dark matter with baryon-phonon coupling}}, \href{https://doi.org/10.1088/1475-7516/2018/09/021}{\emph{JCAP} {\bfseries 09} (2018) 021}, [\href{https://arxiv.org/abs/1711.05748}{{\ttfamily 1711.05748}}].

\bibitem{Landau}
L.~Landau, \emph{{Theory of the Superfluidity of Helium II}}, \href{https://doi.org/10.1103/PhysRev.60.356}{\emph{Phys. Rev.} {\bfseries 60} (1941) 356--358}.

\bibitem{landau1941j}
L.~Landau, \emph{J. eksper. teor. fiz. ussr, 11 (1947)}, {\emph{Phys. Rev} {\bfseries 60} (1941) 356}.

\bibitem{schmitt2015introduction}
A.~Schmitt, \emph{Introduction to superfluidity}, {\emph{Lect. Notes Phys} {\bfseries 888} (2015) }.

\bibitem{Page:2010aw}
D.~Page, M.~Prakash, J.~M. Lattimer and A.~W. Steiner, \emph{{Rapid Cooling of the Neutron Star in Cassiopeia A Triggered by Neutron Superfluidity in Dense Matter}}, \href{https://doi.org/10.1103/PhysRevLett.106.081101}{\emph{Phys. Rev. Lett.} {\bfseries 106} (2011) 081101}, [\href{https://arxiv.org/abs/1011.6142}{{\ttfamily 1011.6142}}].

\bibitem{Rajagopal:2000wf}
K.~Rajagopal and F.~Wilczek, \emph{{The Condensed matter physics of QCD}}, pp.~2061--2151.
\newblock World Scientific, 11, 2000.
\newblock \href{https://arxiv.org/abs/hep-ph/0011333}{{\ttfamily hep-ph/0011333}}.
\newblock 10.1142/9789812810458\_0043.

\bibitem{Witten:1984rs}
E.~Witten, \emph{{Cosmic Separation of Phases}}, \href{https://doi.org/10.1103/PhysRevD.30.272}{\emph{Phys. Rev. D} {\bfseries 30} (1984) 272--285}.

\bibitem{Farhi:1984qu}
E.~Farhi and R.~L. Jaffe, \emph{{Strange Matter}}, \href{https://doi.org/10.1103/PhysRevD.30.2379}{\emph{Phys. Rev. D} {\bfseries 30} (1984) 2379}.

\bibitem{Zhitnitsky:2002qa}
A.~R. Zhitnitsky, \emph{{'Nonbaryonic' dark matter as baryonic color superconductor}}, \href{https://doi.org/10.1088/1475-7516/2003/10/010}{\emph{JCAP} {\bfseries 10} (2003) 010}, [\href{https://arxiv.org/abs/hep-ph/0202161}{{\ttfamily hep-ph/0202161}}].

\bibitem{Alexander:2018fjp}
S.~Alexander, E.~McDonough and D.~N. Spergel, \emph{{Chiral Gravitational Waves and Baryon Superfluid Dark Matter}}, \href{https://doi.org/10.1088/1475-7516/2018/05/003}{\emph{JCAP} {\bfseries 05} (2018) 003}, [\href{https://arxiv.org/abs/1801.07255}{{\ttfamily 1801.07255}}].

\bibitem{Alexander:2020wpm}
S.~Alexander, E.~McDonough and D.~N. Spergel, \emph{{Strongly-interacting ultralight millicharged particles}}, \href{https://doi.org/10.1016/j.physletb.2021.136653}{\emph{Phys. Lett. B} {\bfseries 822} (2021) 136653}, [\href{https://arxiv.org/abs/2011.06589}{{\ttfamily 2011.06589}}].

\bibitem{Ouyed:2023hqe}
R.~Ouyed, D.~Leahy, N.~Koning and P.~Jaikumar, \emph{{Quark Clusters, QCD Vacuum and the Cosmological $^{7}$Li, Dark Matter and Dark Energy Problems}}, \href{https://doi.org/10.3390/universe10030115}{\emph{Universe} {\bfseries 10} (2024) 115}, [\href{https://arxiv.org/abs/2302.06820}{{\ttfamily 2302.06820}}].

\bibitem{Alexander:2024qml}
S.~Alexander, H.~Bernardo and H.~Gilmer, \emph{{A Dark Matter Fermionic Quantum Fluid from Standard Model Dynamics}},  \href{https://arxiv.org/abs/2405.08874}{{\ttfamily 2405.08874}}.

\bibitem{Bernardo:2023ehz}
H.~Bernardo, R.~Brandenberger and A.~Favero, \emph{{Superfluid dark matter flow around cosmic strings}}, \href{https://doi.org/10.1103/PhysRevD.109.123509}{\emph{Phys. Rev. D} {\bfseries 109} (2024) 123509}, [\href{https://arxiv.org/abs/2307.03041}{{\ttfamily 2307.03041}}].

\bibitem{ISRAEL1981}
W.~Israel, \emph{{Covariant superfluid mechanics}}, \href{https://doi.org/https://doi.org/10.1016/0375-9601(81)90169-9}{\emph{Physics Letters A} {\bfseries 86} (1981) 79--81}.

\bibitem{KHALATNIKOV1982}
I.~Khalatnikov and V.~Lebedev, \emph{{Relativistic hydrodynamics of a superfluid liquid}}, \href{https://doi.org/https://doi.org/10.1016/0375-9601(82)90268-7}{\emph{Physics Letters A} {\bfseries 91} (1982) 70--72}.

\bibitem{lebedev1982relativistic}
V.~Lebedev and I.~Khalatnikov, \emph{Relativistic hydrodynamics of a superfluid}, {\emph{Zh. Eksp. Teor. Fiz} {\bfseries 56} (1982) [Sov. Phys. JETP 56, 923 (1982)]}.

\bibitem{ISRAEL1982_2}
W.~Israel, \emph{Equivalence of two theories of relativistic superfluid mechanics}, \href{https://doi.org/https://doi.org/10.1016/0375-9601(82)90298-5}{\emph{Physics Letters A} {\bfseries 92} (1982) 77--78}.

\bibitem{dixon1982thermodynamics}
W.~Dixon, \emph{Thermodynamics of superfluids via relativity}, {\emph{Archive for Rational Mechanics and Analysis} {\bfseries 80} (1982) 159--203}.

\bibitem{Carter:1992gmy}
B.~Carter and I.~M. Khalatnikov, \emph{{Equivalence of convective and potential variational derivations of covariant superfluid dynamics}}, \href{https://doi.org/10.1103/PhysRevD.45.4536}{\emph{Phys. Rev. D} {\bfseries 45} (1992) 4536}.

\bibitem{Carter:1995if}
B.~Carter and D.~Langlois, \emph{{The Equation of state for cool relativistic two constituent superfluid dynamics}}, \href{https://doi.org/10.1103/PhysRevD.51.5855}{\emph{Phys. Rev. D} {\bfseries 51} (1995) 5855--5864}, [\href{https://arxiv.org/abs/hep-th/9507058}{{\ttfamily hep-th/9507058}}].

\bibitem{Son:2000ht}
D.~T. Son, \emph{{Hydrodynamics of relativistic systems with broken continuous symmetries}}, \href{https://doi.org/10.1142/S0217751X01009545}{\emph{Int. J. Mod. Phys. A} {\bfseries 16S1C} (2001) 1284--1286}, [\href{https://arxiv.org/abs/hep-ph/0011246}{{\ttfamily hep-ph/0011246}}].

\bibitem{Nicolis:2011cs}
A.~Nicolis, \emph{{Low-energy effective field theory for finite-temperature relativistic superfluids}},  \href{https://arxiv.org/abs/1108.2513}{{\ttfamily 1108.2513}}.

\bibitem{landau2013fluid}
L.~Landau and E.~Lifshitz, \emph{Fluid Mechanics: Volume 6}.
\newblock No.~v. 6. Elsevier Science, 2013.

\bibitem{London}
F.~London, \emph{{On the Bose-Einstein Condensation}}, \href{https://doi.org/10.1103/PhysRev.54.947}{\emph{Phys. Rev.} {\bfseries 54} (1938) 947--954}.

\bibitem{london1938lambda}
F.~London, \emph{The $\lambda$-phenomenon of liquid helium and the bose-einstein degeneracy}, {\emph{Nature} {\bfseries 141} (1938) 643--644}.

\bibitem{tisza1938transport}
L.~Tisza, \emph{{Transport phenomena in helium II}}, {\emph{Nature} {\bfseries 141} (1938) 913--913}.

\bibitem{Donnelly}
R.~J. Donnelly, \emph{The two-fluid theory and second sound in liquid helium}, \href{https://doi.org/10.1063/1.3248499}{\emph{Physics Today} {\bfseries 62} (2009) 34--39}, [\href{https://arxiv.org/abs/https://doi.org/10.1063/1.3248499}{{\ttfamily https://doi.org/10.1063/1.3248499}}].

\bibitem{BALIBAR2017586}
S.~Balibar, \emph{Laszlo tisza and the two-fluid model of superfluidity}, \href{https://doi.org/https://doi.org/10.1016/j.crhy.2017.10.016}{\emph{Comptes Rendus Physique} {\bfseries 18} (2017) 586--591}.

\bibitem{Dubovsky:2011sj}
S.~Dubovsky, L.~Hui, A.~Nicolis and D.~T. Son, \emph{{Effective field theory for hydrodynamics: thermodynamics, and the derivative expansion}}, \href{https://doi.org/10.1103/PhysRevD.85.085029}{\emph{Phys. Rev. D} {\bfseries 85} (2012) 085029}, [\href{https://arxiv.org/abs/1107.0731}{{\ttfamily 1107.0731}}].

\bibitem{Andersson:2020phh}
N.~Andersson and G.~L. Comer, \emph{{Relativistic fluid dynamics: physics for many different scales}}, \href{https://doi.org/10.1007/s41114-021-00031-6}{\emph{Living Rev. Rel.} {\bfseries 24} (2021) 3}, [\href{https://arxiv.org/abs/2008.12069}{{\ttfamily 2008.12069}}].

\bibitem{2012PhRvD..85h5029D}
S.~{Dubovsky}, L.~{Hui}, A.~{Nicolis} and D.~T. {Son}, \emph{{Effective field theory for hydrodynamics: Thermodynamics, and the derivative expansion}}, \href{https://doi.org/10.1103/PhysRevD.85.085029}{\emph{Physical Review D} {\bfseries 85} (Apr., 2012) 085029}, [\href{https://arxiv.org/abs/1107.0731}{{\ttfamily 1107.0731}}].

\end{thebibliography}\endgroup

\end{document}